\documentclass{ws-procs9x6}

\def\barD{\overline D{}^0}

\def\DDbar{D^0-\overline D{}^0}

\def\D0bar{\overline D{}^0}
\def\K0bar{\overline K{}^0}

\def\3bar{\overline{3}}
\def\sixbar{\overline{6}}

\def\15bar{\overline{15}}
\def\24bar{\overline{24}}
\def\42bar{\overline{42}}
\def\60bar{\overline{60}}
\def\cO{{\it O}}
\def\cal{{\it}}

\def\beq{\begin{equation}}
\def\eeq{\end{equation}}
\def\beqa{\begin{eqnarray}}
\def\eeqa{\end{eqnarray}}
\def\bea{\begin{eqnarray}}
\def\eea{\end{eqnarray}}

\begin{document}

\title{CP-violation and mixing in charmed mesons}

\author{ALEXEY A. PETROV}

\address{Department of Physics and Astronomy\\
Wayne State University\\ 
Detroit, MI 48201, USA\\ 
E-mail: apetrov@physics.wayne.edu}


\maketitle

\abstracts{
The Standard Model contribution to $D^0-{\overline D^0}$ mixing is 
dominated by the contributions of light $s$ and $d$ quarks. 
Neglecting the tiny effects due to $b$ quark, both mass and 
lifetime differences vanish in the limit of $SU(3)_F$ symmetry. 
Thus, the main challenge in the Standard Model calculation of the 
mass and width difference in the $D^0-{\overline D^0}$ system is 
to estimate the size of $SU(3)$ breaking effects. We prove that $D$ 
meson mixing occurs in the Standard Model only at {\it second} 
order in $SU(3)$ violation. We consider the possibility that phase 
space effects may be the dominant source of $SU(3)$ breaking. We 
find that $y=(\Delta \Gamma)/(2\Gamma)$ of the order of one percent 
is natural in the Standard Model, potentially reducing the sensitivity 
to new physics of measurements of $D$ meson mixing. We also discuss 
the possibility of observing lifetime differences and $CP$ violation 
in charmed mesons both at the currently operating and proposed facilities.}

\section{Introduction}

One of the most important motivations for studies of weak decays of
charmed mesons is the possibility of observing a signal from new physics 
which can be separated from the one generated by the Standard Model (SM) 
interactions. The low energy effect of new physics particles can be 
naturally written in terms of a series of local operators of increasing
dimension generating $\Delta C = 1$ (decays) or $\Delta C = 2$ (mixing) 
transitions. For $\DDbar$ mixing these operators,
as well as the one loop Standard Model effects, generate contributions 
to the effective operators that change $D^0$ state into $\barD$ state
leading to the mass eigenstates
\begin{equation} \label{definition1}
| D_{^1_2} \rangle =
p | D^0 \rangle \pm q | \bar D^0 \rangle,
\end{equation}
where the complex parameters $p$ and $q$ are obtained from diagonalizing 
the $D^0-\barD$ mass matrix. The mass and width splittings between these 
eigenstates are parameterized by
\begin{eqnarray} \label{definition}
x \equiv \frac{m_2-m_1}{\Gamma}, ~~
y \equiv \frac{\Gamma_2 - \Gamma_1}{2 \Gamma},
\end{eqnarray}
where $m_{1,2}$ and $\Gamma_{1,2}$ are the masses and widths of
$D_{1,2}$  and the mean width and mass are $\Gamma=(\Gamma_1+\Gamma_2)/2$ 
and $m=(m_1+m_2)/2$. Since $y$ is constructed from the decays of $D$ into 
physical states, it should be dominated by the Standard Model contributions, 
unless new physics significantly modifies $\Delta C=1$ interactions.
On the contrary, $x$ can receive contributions from all energy scales,
so it is usually conjectured that new physics can significantly
modify $x$ leading to the inequality $x\gg y$. As we discuss later, this signal 
for new physics is lost if a relatively large $y$, of the order of a percent, 
is observed. It is known experimentally that $\DDbar$ mixing proceeds extremely 
slow, which in the Standard Model is usually attributed to the absence of superheavy 
quarks destroying GIM cancellations\cite{Petrov:1997ch}.

Another possible manifestation of new physics interactions in the charm
system is associated with the observation of (large) CP-violation. This is due to 
the fact that all quarks that build up the hadronic states in weak 
decays of charm mesons belong to the first two generations. Since $2\times2$ 
Cabbibo quark mixing matrix is real, no CP-violation is possible in the
dominant tree-level diagrams that describe the decay amplitudes. 
In the Standard Model CP-violating amplitudes can be introduced by including 
penguin or box operators induced by virtual $b$-quarks. However, their 
contributions are strongly suppressed by the small combination of 
CKM matrix elements $V_{cb}V^*_{ub}$. It is thus widely believed that the 
observation of (large) CP violation in charm decays or mixing would be an 
unambiguous sign for new physics. This fact makes charm decays a valuable 
tool in searching for new physics, since the statistics available in charm 
physics experiment is usually quite large.

As in B-physics, CP-violating contributions in charm can be generally classified 
by three different categories:
(I) CP violation in the decay amplitudes. This type of CP violation 
occurs when the absolute value of the decay amplitude for $D$ to decay to a 
final state $f$ ($A_f$) is different from the one of corresponding CP-conjugated 
amplitude (``direct CP-violation'');
(II) CP violation in $\DDbar$ mixing matrix. This type of CP violation
is manifest when 
$R_m^2=\left|p/q\right|^2=(2 M_{12}-i \Gamma_{12})/(2 M_{12}^*-i \Gamma_{12}^*) \neq 1$; 
and 
(III) CP violation in the interference of decays with and without mixing.
This type of CP violation is possible for a subset of final states to which
both $D^0$ and $\barD$ can decay. 

For a given final state $f$, CP violating contributions can be summarized in the
parameter 
\beq
\lambda_f = \frac{q}{p} \frac{{\overline A}_f}{A_f}=
R_m e^{i(\phi+\delta)}\left| \frac{{\overline A}_f}{A_f}\right|,
\eeq
where $A_f$ and ${\overline A}_f$ are the amplitudes for $D^0 \to f$ and 
$\barD \to f$ transitions respectively and $\delta$ is the strong phase 
difference between $A_f$ and ${\overline A}_f$. Here $\phi$ represents the
convention-independent weak phase difference between the ratio of 
decay amplitudes and the mixing matrix.

\section{Present and perspective experimental constraints}

Presently, experimental information about the $\DDbar$ mixing parameters 
$x$ and $y$ comes from the time-dependent analyses that can roughly be divided
into two categories. First, more traditional studies look at the time
dependence of $D \to f$ decays, where $f$ is the final state that can be
used to tag the flavor of the decayed meson. The most popular is the
non-leptonic doubly Cabibbo suppressed decay (DCSD) $D^0 \to K^+ \pi^-$.
Time-dependent studies allow one to separate the DCSD from the mixing 
contribution $D^0 \to \D0bar \to K^+ \pi^-$,
\begin{eqnarray}\label{Kpi}
\Gamma[D^0(t) \to K^+ \pi^-]
=e^{-\Gamma t}|A_{K^-\pi^+}|^2 \qquad\qquad\qquad\qquad\qquad\qquad\qquad
\nonumber \\
\times ~\left[
R+\sqrt{R}R_m(y'\cos\phi-x'\sin\phi)\Gamma t
+\frac{R_m^2}{4}(y^2+x^2)(\Gamma t)^2
\right],
\end{eqnarray}
where $R$ is the ratio of DCS and Cabibbo favored (CF) decay rates. 
Since $x$ and $y$ are small, the best constraint comes from the linear terms 
in $t$ that are also {\it linear} in $x$ and $y$.
A direct extraction of $x$ and $y$ from Eq.~(\ref{Kpi}) is not possible due 
to unknown relative strong phase $\delta$ of DCS and CF 
amplitudes\cite{Falk:1999ts}, 
as $x'=x\cos\delta+y\sin\delta$, $y'=y\cos\delta-x\sin\delta$. This phase can be 
measured independently\cite{GGR}. The corresponding formula can also be 
written\cite{Bergmann:2000id} for $\barD$ decay with $x' \to -x'$ and $R_m \to R_m^{-1}$.

Second, $D^0$ mixing can be measured by comparing the lifetimes 
extracted from the analysis of $D$ decays into the CP-even and CP-odd 
final states. This study is also sensitive to a {\it linear} function of 
$y$ via
\beq
\frac{\tau(D \to K^-\pi^+)}{\tau(D \to K^+K^-)}-1=
y \cos \phi - x \sin \phi \left[\frac{R_m^2-1}{2}\right].
\eeq
Time-integrated studies of the semileptonic transitions are sensitive
to the {\it quadratic} form $x^2+y^2$ and at the moment are not 
competitive with the analyses discussed above. 

The construction of a new tau-charm factory at Cornell (CLEO-c) will introduce 
new {\it time-independent} methods that are sensitive to a linear function of 
$y$. In particular, one can use the fact that heavy meson 
pairs produced in the decays of heavy quarkonium resonances have the
useful property that the two mesons are in the CP-correlated states\cite{AtwoodPetrov}.
By tagging one of the mesons as a CP eigenstate, a lifetime difference 
may be determined by measuring the leptonic branching ratio of the other meson.
The initial $D^0 \barD$ state is prepared as 
\begin{eqnarray}
|D \barD \rangle_L = \frac{1}{\sqrt{2}} 
\left \{
| D^0 (k_1)\barD (k_2) \rangle +
(-1)^L | D^0 (k_2)\barD (k_1) \rangle
\right \},
\end{eqnarray}
where $L$ is the relative angular momentum of two $D$ mesons. There are several 
possible resonances at which CLEO-c will be running, for example $\psi(3770)$ 
where $L = 1$ and the initial state is antisymmetric, or $\psi(4114)$ where the 
initial  $D^0 \barD$ state can be symmetric due to emission of additional pion or
photon in the decay. In this scenario, the CP quantum numbers of the $D(k_2)$ can 
be determined. The semileptonic {\it width} of this meson should be independent of 
the CP quantum number since it is flavor specific. It follows that the semileptonic 
{\it branching ratio} of $D(k_2)$ will be inversely proportional to the total 
width of that meson. Since we know whether $D(k_2)$ is tagged as a (CP-eigenstate) 
$D_+$ or and $D_-$ from the decay of $D(k_1)$ to $S_\sigma$, we can easily determine 
$y$ in terms of the semileptonic branching ratios of $D_\pm$. This can be expressed 
simply by introducing the ratio
\beq \label{DefCor}
R^L_\sigma=
\frac{\Gamma[\psi_L \to (H \to S_\sigma)(H \to X l^\pm \nu )]}{
\Gamma[\psi_L \to (H \to S_\sigma)(H \to X)]~Br(H^0 \to X l \nu)},
\eeq
where $X$ in $H \to X$ stands for an inclusive set of all
final states. A deviation from $R^L_\sigma=1$ implies a
lifetime difference. Keeping only the leading (linear) contributions
due to mixing, $y$ can be extracted from this experimentally obtained 
quantity,
\begin{eqnarray}
y\cos\phi=
(-1)^L {\sigma}
{R^L_\sigma-1\over R^L_\sigma}
\label{y-cos-phi}.
\end{eqnarray}

The current experimental upper bounds on $x$ and $y$ are on the order of 
a few times $10^{-2}$, and are expected to improve significantly in the coming
years.  To regard a future discovery of nonzero $x$ or $y$ as a signal for new 
physics, we would need high confidence that the Standard Model predictions lie
well below the present limits.  As was recently shown\cite{Falk:2001hx}, 
in the Standard Model $x$ and $y$ are generated only at second order in $SU(3)$ 
breaking, 
\beq
x\,,\, y \sim \sin^2\theta_C \times [SU(3) \mbox{ breaking}]^2\,,
\eeq
where $\theta_C$ is the Cabibbo angle.  Therefore, predicting the
Standard Model values of $x$ and $y$ depends crucially on estimating the 
size of $SU(3)$ breaking.  Although $y$ is expected to be determined
by the Standard Model processes, its value nevertheless affects significantly 
the sensitivity to new physics of experimental analyses of $D$ 
mixing\cite{Bergmann:2000id}.

Theoretical calculations of $x$ and $y$, as will be discussed later, are 
quite uncertain, and the values near the current experimental bounds cannot be 
ruled out. Therefore, it will be difficult to find a clear indication of 
physics beyond the Standard Model in $\DDbar$ mixing measurements alone.
The only robust potential signal of new physics in charm system at this stage 
is CP violation.

CP violation in $D$ decays and mixing can be searched for by a variety of 
methods. For instance, time-dependent decay widths for $D \to K \pi$ are 
sensitive to CP violation in mixing (see Eq.(\ref{Kpi})). Provided that 
the $x$ and $y$ are comparable to experimental sensitivities, a combined 
analysis of $D \to K \pi$ and $D \to KK$ can yield interesting 
constraints on CP-violating parameters\cite{Bergmann:2000id}.

Most of the techniques that are sensitive to CP violation make use of the
decay asymmetry,
\begin{eqnarray}\label{Acp}
A_{CP}(f)=\frac{\Gamma(D \to f)-\Gamma({\overline D} \to {\overline f})}{
\Gamma(D \to f)+\Gamma({\overline D} \to {\overline f})}=
\frac{1-\left|{\overline A}_{\overline f}/A_f\right|^2}{1+
\left|{\overline A}_{\overline f}/A_f\right|^2}.
\end{eqnarray}
Most of the properties of Eq.(\ref{Acp}), such as dependence on the
strong final state phases, are similar to the ones in B-physics\cite{BigiSandaBook}.
Current experimental bounds from various experiments, all consistent
with zero within experimental uncertainties, can be found in\cite{Pedrini:2000ge}.

Other interesting signals of $CP$-violation that are being 
discussed in connection with tau-charm factory measurements are
the ones that are using quantum coherence of the initial state.
An example of this type of signal is a decay $(D^0 \barD) \to f_1 f_2$ at 
$\psi(3770)$ with $f_1$ and $f_2$ being the different final CP-eigenstates
with $CP|f_1\rangle=CP|f_2\rangle$. This type of signals are very easy to detect 
experimentally. It is easy to compute this CP-violating decay rate for the final states
$f_1$ and $f_2$
\begin{eqnarray} \label{CPrate}
\Gamma_{f_1 f_2}=
\frac{\left(2+x^2-y^2\right) \left|\lambda_{f_1}-\lambda_{f_2}\right|^2
+\left(x^2+y^2\right)\left|1-\lambda_{f_1} \lambda_{f_2}\right|^2 }
{2 R_m^2 (1+x^2)(1-y^2)} ~\Gamma_{f_1} \Gamma_{f_2}
\end{eqnarray}
The result of Eq.~(\ref{CPrate}) represents a generalization of the formula 
given in Ref.~\cite{Bigi:1986dp}. It is clear that both terms in the numerator 
of Eq.~(\ref{CPrate}) receive contributions from CP-violation of the type I 
and III, while the second term is also sensitive to CP-violation of the
type II. Moreover, for a large set of the final states the first term would be 
additionally suppressed by $SU(3)$ symmetry. For instance, $\lambda_{\pi\pi}=\lambda_{KK}$ 
in the $SU(3)$ symmetry limit. It is easy to see that only the second term survives if only 
CP violation in the mixing matrix is retained,
$\Gamma_{f_1 f_2} \propto \left|1-R_m^2\right|^2 \propto A_m^2$. This expression is of 
the {\it second} order in CP-violating parameters. As it follows from the existing experimental 
constraints on rate asymmetries, CP-violating phases are quite small in charm system, regardless 
of whether they are produced by the Standard Model mechanisms or by some new physics 
contributions. In that respect, it looks unlikely that the SM signals of CP violation 
would be observed at CLEO-c with this observable.

While the searches for direct CP violation via the asymmetry of Eq.~(\ref{Acp}) can be
done with the charged D-mesons (which are self-tagging), investigations of the other two 
types of CP-violation require flavor tagging of the initial state. This severely
cuts the available dataset. It is therefore interesting to look for signals of CP violation
that do not require identification of the initial state. One possible CP-violating 
signal involves the observable obtained by summing over the initial 
states, $\sum \Gamma_i=\Gamma_i+{\overline \Gamma}_i$ for $i=f,{\overline f}$.
A CP-odd observable that can be formed out of $\sum \Gamma_i$ is an asymmetry
\begin{equation} \label{TotAsym}
A_{CP}^U=\frac{\sum \Gamma_f - \sum \Gamma_{\overline f}}{\sum \Gamma_f + 
\sum \Gamma_{\overline f}}.
\end{equation}
Note that this asymmetry does not require quantum coherence of the initial state 
and therefore is accessible in any D-physics experiment. The final states must be 
chosen such that $A_{CP}^U$ is not trivially zero. As we shall see below, 
decays of $D$ into the final states that are CP-eigenstates 
would result in zero asymmetry, while the final states like $K^+ K^{*-}$ 
or $K_S \pi^+ \pi^-$ would not. A non-zero value of $A_{CP}^U$ in 
Eq.~(\ref{TotAsym}) can be generated by both direct and indirect 
CP-violating contributions. These can be separated by appropriately 
choosing the final states. For example, indirect CP violating amplitudes are
tightly constrained in the decays dominated by the Cabibbo-favored 
tree level amplitudes, while singly Cabibbo suppressed amplitudes 
also receive contributions from direct CP violating amplitudes. 
Neglecting small CP-violation in the mixing matrix ($R_m \to 1$) 
one obtains,
\begin{eqnarray}
A_{CP}^U &=& \frac{\Gamma_f-{\overline \Gamma}_{\overline f}-
\Gamma_{\overline f}+{\overline \Gamma}_f}
{\Gamma_f + \Gamma_{\overline f} + {\overline \Gamma}_f
+{\overline \Gamma}_{\overline f}}
+ \frac{2 y}{\Gamma_f + \Gamma_{\overline f} + {\overline \Gamma}_f
+{\overline \Gamma}_{\overline f}}
\nonumber \\
&\times&\left[
\cos\phi \left(Re {\overline A}_f^*  {\overline A}_{\overline f}
-Re {A}_f^*  {A}_{\overline f}\right)
+ \sin\phi \left( Im {\overline A}_f  {\overline A}_{\overline f}^*
+Im {A}_f^* {A}_{\overline f}\right) 
\right].~~~
\end{eqnarray}
It is easy to see that, as promised, this asymmetry vanishes for the 
final states that are CP-eigenstates, as 
${\Gamma }_{f}={\Gamma }_{\overline f}$ and 
${\Gamma }_{f} - {\overline \Gamma}_{\overline f} =
{\Gamma}_{\overline f} - {\overline \Gamma }_f$.

\section{Theoretical expectations for mixing parameters}

Theoretical predictions of $x$ and $y$ within and beyond
the Standard Model span several orders of magnitude\cite{Nelson:1999fg}.
Roughly, there are two approaches, neither of which give very reliable
results because $m_c$ is in some sense intermediate between heavy and
light.  The ``inclusive'' approach is based on the operator
product expansion (OPE).  In the $m_c \gg \Lambda$ limit, where
$\Lambda$ is a scale characteristic of the strong interactions, $\Delta
M$ and $\Delta\Gamma$ can be expanded in terms of matrix elements of local
operators\cite{Inclusive}.  Such calculations yield $x,y < 10^{-3}$.  
The use of the OPE relies on local quark-hadron duality, 
and on $\Lambda/m_c$ being small enough to allow a truncation of the series
after the first few terms.  The charm mass may not be large enough for these 
to be good approximations, especially for nonleptonic $D$ decays.
An observation of $y$ of order $10^{-2}$ could be ascribed to a
breakdown of the OPE or of duality,  but such a large
value of $y$ is certainly not a generic prediction of OPE analyses.
The ``exclusive'' approach sums over intermediate hadronic
states, which may be modeled or fit to experimental data\cite{Exclusive}.
Since there are cancellations between states within a given $SU(3)$
multiplet, one needs to know the contribution of each state with high 
precision. However, the $D$ is not light enough that its decays are dominated
by a few final states.  In the absence of sufficiently precise data on many decay 
rates and on strong phases, one is forced to use some assumptions.  While most 
studies find $x,y < 10^{-3}$, Refs.\cite{Exclusive} obtain $x$ and 
$y$ at the $10^{-2}$ level by arguing that $SU(3)$ violation is of order
unity, but the source of the large $SU(3)$ breaking is not made explicit.

In what follows we first prove that $D^0-\D0bar$ mixing arises only at 
{\it second} order in $SU(3)$ breaking effects.  The proof is valid when
$SU(3)$ violation enters perturbatively. This would not be so, for
example, if $D$ transitions were dominated by a single narrow 
resonance close to threshold\cite{Falk:2001hx,Golowich:1998pz}. Then we argue that 
reorganization of ``exclusive'' calculation by explicitly building 
$SU(3)$ cancellations into the analysis naturally leads to values of 
$y \sim 1\%$ if only one source of $SU(3)$ breaking (phase space) is taken 
into account.

The quantities $M_{12}$ and $\Gamma_{12}$ which determine $x$ and $y$
depend on matrix elements 
$\langle\D0bar|\, {\cal H}_w {\cal H}_w\, |D^0\rangle\,$,
where ${\cal H}_w$ denote the $\Delta C=-1$ part of the weak Hamiltonian.  
Let $D$ be the field operator that creates a $D^0$ meson and annihilates a
$\D0bar$.  Then the matrix element, whose $SU(3)$ flavor group theory properties
we will study, may be written as
\beq\label{melm}
      \langle 0|\, D\, {\cal H}_w {\cal H}_w \,D\, |0 \rangle\,.
\eeq
Since the operator $D$ is of the form $\bar cu$, it transforms in the
fundamental representation of $SU(3)$, which we will represent with a
lower index, $D_i$.  We use a convention in which the correspondence between 
matrix indexes and quark flavors is $(1,2,3)=(u,d,s)$.  The only nonzero 
element of $D_i$ is $D_1=1$.  The $\Delta C=-1$
part of the weak Hamiltonian has the flavor structure $(\bar q_ic)(\bar
q_jq_k)$, so its matrix representation is written with a fundamental
index and two antifundamentals, $H^{ij}_k$.  This operator is a sum of irreps
contained in the product $3 \times \3bar \times \3bar =
\15bar + 6 + \3bar + \3bar$.  In the limit in which the third generation is
neglected, $H^{ij}_k$ is traceless, so only the $\15bar$ 
and 6 representations appear.  That is, the
$\Delta C=-1$ part of ${\cal H}_w$ may be decomposed as ${1\over2} 
(\cO_{\15bar} + \cO_6)$,
where
\beqa
\cO_{\15bar} &=& (\bar sc)(\bar ud) + (\bar uc)(\bar sd)
    + s_1(\bar dc)(\bar ud) + s_1(\bar uc)(\bar dd)\nonumber\\
&&{} - s_1(\bar sc)(\bar us) - s_1(\bar uc)(\bar ss)
    - s_1^2(\bar dc)(\bar us) - s_1^2(\bar uc)(\bar ds) \,, \nonumber\\
\cO_6 &=& (\bar sc)(\bar ud) - (\bar uc)(\bar sd)
    + s_1(\bar dc)(\bar ud) - s_1(\bar uc)(\bar dd)\nonumber\\
&&{} - s_1(\bar sc)(\bar us) + s_1(\bar uc)(\bar ss)
    - s_1^2(\bar dc)(\bar us) + s_1^2(\bar uc)(\bar ds) \,,
\eeqa
and $s_1=\sin\theta_C$.  The matrix representations
$H(\15bar)^{ij}_k$ and $H(6)^{ij}_k$ have nonzero elements
\begin{equation}
\begin{tabular}{rll}
$H(\15bar)^{ij}_k:\qquad$
    &  $H^{13}_2 = H^{31}_2=1$\,,  &  $H^{12}_2 = H^{21}_2 = s_1$\,,\\
&  $H^{13}_3 = H^{31}_3 = -s_1$\,,  &  $H^{12}_3 =
H^{21}_3=-s_1^2$\,,\\[4pt]
$H(6)^{ij}_k:\qquad$
    &  $H^{13}_2 = -H^{31}_2=1$\,,  &  $H^{12}_2 = -H^{21}_2 = s_1$\,,\\
&  $H^{13}_3 = -H^{31}_3 = -s_1$\,,$\qquad$
    &  $H^{12}_3 = -H^{21}_3 = -s_1^2$\,.
\end{tabular}
\end{equation}
We introduce $SU(3)$ breaking through the quark mass operator ${\cal
M}$, whose matrix representation is $M^i_j={\rm diag}(m_u,m_d,m_s)$
as being in the adjoint representation to induce $SU(3)$ violating effects.  
We set $m_u=m_d=0$ and let $m_s\ne0$ be the only $SU(3)$ violating parameter.  
All nonzero matrix elements built out of $D_i$, $H^{ij}_k$ and $M^i_j$ must be 
$SU(3)$ singlets.

We now prove that $D^0-\D0bar$ mixing arises only at second order in
$SU(3)$ violation, by which we mean second order in $m_s$.  First, we note that
the pair of $D$ operators is symmetric, and so the product $D_iD_j$
transforms as a 6 under $SU(3)$.  Second, the pair of ${\cal H}_w$'s is also symmetric, 
and the product $H^{ij}_kH^{lm}_n$ is in one of the reps which
appears in the product
\beqa
\left[ (\15bar+6)\times(\15bar+6) \right]_S =
    (\15bar\times\15bar)_S +(\15bar\times 6)+(6\times 6)_S ~~~~~~~~~\\*
= (\60bar+\24bar+15+15'+\sixbar) + (42+24+15+\sixbar+3)
    + (15'+\sixbar)\,. \nonumber
\eeqa
A direct computation shows that only three of these
representations actually appear in the decomposition of ${\cal H}_w{\cal H}_w$.  They 
are the $\60bar$, the 42, and the $15'$ (actually twice, but with the same nonzero elements
both times).  So we have product operators of the form (the subscript denotes the 
representation of $SU(3)$)
\beqa
    DD = {\cal D}_6\,, ~~~~
    {\cal H}_w {\cal H}_w = \cO_{\60bar}+\cO_{42}+\cO_{15'}\,.
\eeqa
Since there is no $\sixbar$ in the decomposition of ${\cal H}_w{\cal 
H}_w$, there is no $SU(3)$ singlet which can be made with ${\cal D}_6$,  
and no $SU(3)$ invariant matrix element of the form (\ref{melm}) can be formed.  
This is the well known result that $D^0-\D0bar$ mixing is 
{\it prohibited by $SU(3)$ symmetry}.
Now consider a single insertion of the $SU(3)$ violating spurion ${\cal M}$.
The combination ${\cal D}_6{\cal M}$ transforms as $6\times
8=24+\15bar+6+\3bar$. There is still no invariant to be made
with ${\cal H}_w{\cal H}_w$, thus $D^0-\D0bar$ mixing is {\it not 
induced at first order in $SU(3)$ breaking}.
With two insertions of ${\cal M}$, it becomes possible to make an $SU(3)$
invariant.  The decomposition of ${\cal D}{\cal M}{\cal M}$ is
\beqa
6\times(8\times 8)_S &=& 6\times(27+8+1) \\
    &=& (60+\42bar+24+\15bar+\15bar'+6) + (24+\15bar+6+\3bar) + 6\,.
\nonumber
\eeqa
There are three elements of the $6\times 27$ part which can give invariants
with  ${\cal H}_w{\cal H}_w$.  Each invariant yields a contribution to 
$D^0-\D0bar$ mixing proportional to $s_1^2m_s^2$. Thus, $\DDbar$ mixing arises 
only at {\it second order} in the $SU(3)$ violating parameter $m_s$.

We now turn to the contributions to $y$ from on-shell final states, which result
from every common decay product of $D^0$ and $\D0bar$.  In the $SU(3)$ limit, 
these contributions cancel when one sums over complete $SU(3)$ multiplets in 
the final state.  The cancellations
depend on $SU(3)$ symmetry both in the decay matrix elements and in the
final state phase space.  While there are $SU(3)$ violating
corrections to both of these, it is difficult to compute the $SU(3)$
violation in the matrix elements in a model independent manner. Yet, with 
some mild assumptions about the momentum dependence of the matrix elements, the 
$SU(3)$ violation in the phase space depends only on the final particle masses and 
can be computed. We estimate the contributions to $y$ solely from 
$SU(3)$ violation in the phase space. We find that this source of $SU(3)$ 
violation can generate $y$ of the order of a few percent.

The mixing parameter $y$ may be written in terms of the matrix elements
for common final states for $D^0$ and $\D0bar$ decays,
\beq
y = {1\over\Gamma} \sum_n \int [{\rm P.S.}]_n\,
    \langle \D0bar|\,{\cal H}_w\,|n \rangle \langle n|\,{\cal H}_w\,|D^0
\rangle\,,
\eeq
where the sum is over distinct final states $n$ and the integral is over
the phase space for state $n$.  Let us now perform the phase space integrals
and restrict the sum to final states $F$ which  transform within a
single $SU(3)$ multiplet $R$.  The result is a  contribution to $y$ of
the form
\beq
    {1\over\Gamma}\, \langle\D0bar|\,{\cal H}_w
     \bigg\{ \eta_{CP}(F_R)\sum_{n\in  F_R}
    |n\rangle \rho_n\langle n| \bigg\} {\cal H}_w\,|D^0\rangle\,,
\eeq
where $\rho_n$ is the phase space available to the state $n$, 
$\eta_{CP}=\pm1$~\cite{Falk:2001hx}.  In the
$SU(3)$ limit, all the $\rho_n$ are the same for $n\in F_R$, and the quantity 
in braces above is an $SU(3)$ singlet.  Since the $\rho_n$ depend only on the 
known masses of the particles in the state $n$, incorporating the true values
of $\rho_n$ in the sum is a calculable source of $SU(3)$ breaking.

This method does not lead directly to a calculable contribution to $y$,
because
the matrix elements $\langle n|{\cal H}_w|D^0\rangle$ and
$\langle\D0bar|{\cal H}_w|n\rangle$
are not known.  However, $CP$ symmetry, which in the Standard Model and
almost all scenarios of new physics is to an excellent approximation
conserved in $D$ decays, relates $\langle\D0bar|{\cal H}_w|n\rangle$ to
$\langle
D^0|{\cal H}_w|\overline{n}\rangle$. Since $|n\rangle$ and
$|\overline{n}\rangle$ are
in a common $SU(3)$ multiplet,  they are determined by a single effective
Hamiltonian. Hence the ratio
\beqa\label{yfr}
   y_{F,R} &=& {\sum_{n\in F_R} \langle\D0bar|\,{\cal H}_w|n\rangle \rho_n
    \langle n|{\cal H}_w\,|D^0\rangle \over
    \sum_{n\in F_R} \langle D^0|\,{\cal H}_w |n\rangle \rho_n
    \langle n|{\cal H}_w\,|D^0\rangle} \nonumber \\
   &=& {\sum_{n\in F_R} \langle\D0bar|\,{\cal H}_w|n\rangle \rho_n
    \langle n|{\cal H}_w\,|D^0\rangle \over \sum_{n\in F_R}
   \Gamma(D^0\to n)} 
\eeqa
is calculable, and represents the value which $y$ would take if elements
of $F_R$ were the only channel open for $D^0$ decay.  To get a true
contribution
to $y$, one must scale $y_{F,R}$ to the total branching  ratio to all the
states in $F_R$.  This is not trivial, since a given physical final state
typically decomposes into a sum over more than one multiplet $F_R$.  The
numerator of $y_{F,R}$ is of order $s_1^2$ while the  denominator is of
order 1, so with large $SU(3)$ breaking in the phase space the natural size 
of $y_{F,R}$ is 5\%.
Indeed, there are other $SU(3)$ violating effects, such as in matrix elements 
and final state interaction phases.  Here we assume that
there is no cancellation with other sources of $SU(3)$ breaking, or
between the various multiplets which occur in $D$ decay, that would
reduce our result for $y$ by an order of magnitude.
This is equivalent to assuming that the $D$ meson is not
heavy enough for duality to enforce such cancellations. Performing the 
computations of $y_{F,R}$, we see\cite{Falk:2001hx} that effects at the level 
of a few percent are quite generic.  
Our results are  summarized in Table~\ref{ytwobody}. Then, $y$ can be
formally constructed from the individual $y_{F,R}$ by
weighting them by their $D^0$ branching ratios,
\beq\label{ycombine}
    y = {1\over\Gamma} \sum_{F,R}\, y_{F,R}
    \bigg[\sum_{n\in F_R}\Gamma(D^0\to n)\bigg]\,.
\eeq
However, the data on $D$ decays are neither abundant nor precise enough
to disentangle the decays to the various $SU(3)$ multiplets, especially
for the three- and four-body final states.  Nor have we computed  $y_{F,R}$ for
all or even most of the available representations.   Instead, we can only
estimate individual contributions to $y$ by  assuming that the representations
for which we know $y_{F,R}$ to be  typical for final states with a given
multiplicity, and then to scale to  the total branching ratio to those final states.
The total branching  ratios of $D^0$ to two-, three- and four-body final
states can be extracted from the Review of Particle Physics\cite{PDG}. Rounding to the 
nearest 5\% to emphasize 
the uncertainties in these numbers, we conclude that the branching fractions for 
$PP$, $(VV)_{\mbox{$s$-wave}}$, $(VV)_{\mbox{$d$-wave}}$ and $3P$ approximately 
amount to 5\%, while the branching ratios for $PV$ and $4P$ are of the order 
of 10\%\cite{Falk:2001hx}.

\begin{table}
\begin{center}
\begin{tabular}{|@{~~~}lc|c|c|} \hline
\multicolumn{2}{|c|}{~~Final state representation~~~}  &
     ~~~$y_{F,R}/s_1^2$~~~ & ~~~$y_{F,R}\ (\%)$~~~  \\ \hline\hline
    $PP$  &  $8$  &  $-0.0038$ & $-0.018$  \\
    &  $27$  &  $-0.00071$  & $-0.0034$ \\ \hline
    $PV$  &  $8_A$  &  $0.032$ & $0.15$\\
    &  $8_S$  &  $0.031$  & $0.15$ \\
    &  $10$  &  $0.020$ & $0.10$ \\
    &  $\overline{10}$  &  $0.016$ & $0.08$ \\
    &  $27$  &  $0.04$  & $0.19$\\ \hline
    $(VV)_{\mbox{$s$-wave}}$  &  $8$  &  $-0.081$ & $-0.39$ \\
    &  $27$  &  $-0.061$ & $-0.30$\\
    $(VV)_{\mbox{$p$-wave}}$  &  $8$  &  $-0.10$ & $ -0.48$\\
    &  $27$  &  $-0.14$ & $-0.70$ \\
    $(VV)_{\mbox{$d$-wave}}$  &  $8$  &  $0.51$ & $2.5$ \\
    &  $27$  &  $0.57$  & $2.8$\\ \hline
$(3P)_{\mbox{$s$-wave}}$        &  $8$  &  $-0.48$  & $-2.3$\\
    &  $27$  &  $-0.11$  & $-0.54$ \\
$(3P)_{\mbox{$p$-wave}}$        &  $8$  &  $-1.13$  & $-5.5$ \\
    &  $27$  &  $-0.07$   & $-0.36$  \\
$(3P)_{\mbox{form-factor}}$     &  $8$  &  $-0.44$  & $-2.1$\\
    &  $27$  &  $-0.13$ & $-0.64$ \\ \hline
$4P$  &  $8$  &  $3.3$ & $16$  \\
    &  $27$  &  $2.2$  & $11$ \\
    &  $27'$  &  $1.9$ & $9.2$ \\ \hline
\end{tabular} \vspace{4pt}
\caption{Values of $y_{F,R}$ for some two-, three-, and four-body final states.}
\label{ytwobody}
\end{center}
\end{table}

We observe that there are terms in Eq.~(\ref{ycombine}), like nonresonant $4P$, 
which could make contributions  to $y$ at the level of a percent or larger.  
There, the rest masses of the final state particles take up most of the available 
energy, so phase space differences are very important. One can see
that $y$  on the order  of a few percent is completely natural, and that anything 
an order of magnitude smaller would require significant  cancellations which do not
appear naturally in this framework.  Cancellations would be expected only if
they were enforced by the OPE, or if the charm quark were heavy
enough that the ``inclusive'' approach were applicable. The hypothesis
underlying the present analysis is that this is not the case.

\section{Conclusions}

We proved that if $SU(3)$ violation may be treated perturbatively,
then $D^0 - \D0bar$ mixing in the Standard Model is generated only at
second order in $SU(3)$ breaking effects.
Within the exclusive approach, we identified an $SU(3)$ breaking effect,
$SU(3)$ violation in final state phase space, which can be calculated
with minimal model dependence.  We found that phase space effects
alone provide enough $SU(3)$ violation to induce $y\sim10^{-2}$.
Large effects in $y$ appear for decays close to $D$ threshold, where
an analytic expansion in $SU(3)$ violation is no longer possible.

Indeed, some degree of cancellation is possible between different
multiplets, as would be expected in the $m_c\to \infty$ limit, or between
$SU(3)$ breaking in phase space and in matrix elements.  It is not known
how effective these cancellations are, and the most reasonable assumption 
in light of our analysis is that they are not significant enough to result 
in an order of magnitude suppression of $y$, as they are not enforced by any
symmetry arguments. Therefore, any future discovery of a $D$ meson width 
difference should not by itself be interpreted as an indication of the 
breakdown of the Standard Model.

At this stage the only robust potential signal of new physics in charm system 
is CP violation. We discussed several possible experimental observables that are
sensitive to CP violation.

\section*{Acknowledgments}
It is my pleasure to thank D. Atwood, S. Bergmann, E. Golowich, Y. Grossman, A. Falk, 
Z. Ligeti, and Y. Nir for collaborations on the related projects. I would like to 
thank the organizers for the invitation to the wonderfully organized Arkadyfest 
workshop.

\end{document}